# Subwavelength focusing of azimuthally polarized beams with vortical phase in dielectrics by using an ultra-thin lens

Kun Huang,[1] Huapeng Ye,[1] Hong Liu,[2] Jinghua Teng,[2] Swee Ping Yeo,[1] and Cheng-Wei Qiu[1,*]

[1]*Department of Electrical and Computer Engineering, National University of Singapore, 4 Engineering Drive 3, Singapore 117576*
[2]*Institute of Materials Research and Engineering, Agency for Science, Technology and Research (A*STAR), 3 Research Link, Singapore 117602*
*\*Corresponding author: eleqc@nus.edu.sg*

We demonstrate that a planar and ultrathin binary lens can focus an azimuthally polarized beam with vortical phase (APV) to a subwavelength spot of transverse polarization. The results elaborates that, in the multi-layer medium, this focused spot, which is beyond the Rayleigh diffraction limitation, can be well maintained for several wavelengths after travelling through the dielectric interfaces, which is not attainable by using other vector beams (i.e., radially, linearly and circularly polarized beams) as the illuminating light. This compact optical system can be valuable in data writing and defect identification of wafer or silicon chips, owing to the enhanced polarized focusing through interfaces. It also enables to be highly integrated with traditional microscopy for the far-field super-resolution imaging, surface scanning and detection, and subwavelength focusing, owing to the enhanced focusing performance (reduced width and extended length) as well as the planarized configuration of the ultrathin lens.

The polarization of light has shown its strong effect in tight focusing [1, 2], high-resolution imaging [3], controlling the nonlinear process [4 and quantum communication [5]. The beams with novel polarizations, i.e. cylindrical vector beams [6, 7] and Poincare beams [8], have received much attention in the past few years. In particular, the radially and azimuthally polarized beams are well known for unique properties in focusing and imaging [1]. Being focused by a high numerical-aperture (NA) spherical lens, the radially polarized (RP) beam has a subwavelength focusing spot at the focal plane because of its strong longitudinal component [1, 9]. Due to its purely transverse electric field in the focal region, the azimuthally polarized beam with vortical phase (APV) has more advantages than the RP beam in focusing light into a small spot by a high NA spherical lens in homogenous or multi-layer medium [10, 11, 12].

However, the requirement on compactness and miniaturization of elements in integrated optics [13] impose big challenges to the traditional optical elements in terms of size, planarity, volume and capacity in manipulating the amplitude, phase and polarization of light simultaneously [14,15]. For example, an ultrathin metallic lens with alternating annuli of metal and air apertures enables a subwavelength focusing spot in air with longitudinal polarization[16] beyond the diffraction limit, making it a competitive candidate as focusing lens in integrated optics. Nevertheless, this longitudinally polarized beam is discontinuous at the interface of two neighboring media and enlarges the focusing spot in the high-NA medium. This implies that the subwavelength focused spot with longitudinal polarization is limited in application such as imaging the silicon integrated circuit [17]. In order to solve this problem, we here propose that APV beam, whose electric field is $\mathbf{E}(r,\varphi)=P(r)e^{i\varphi}\mathbf{e}_\varphi$ with the amplitude factor $P(r)$ and the polar coordinates $r$ and $\varphi$, is used as the illuminating light of binary lens. Our results unambiguously show that APV beam is superior in achieving a subwavelength focused spot regardless of the presence of the medium interface, and can maintain this subwavelength focusing spot for several wavelengths even after passing through interface. Moreover, we also display that APV beam has a super-resolution focused spot in a homogenous medium when focused by a binary lens with a large range of NA.

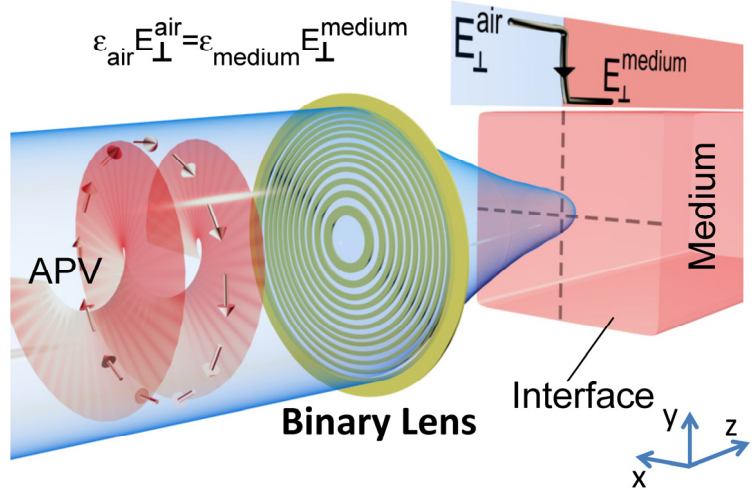

**Figure 1** (color online)Tight focusing of APV beam by a binary lens in the presence of two-layer media. The phase and polarization of APV are shown by the helical structure and the space-variant arrows, respectively. At the interface between air and medium, two normal components ($E_\perp^{air}$ and $E_\perp^{medium}$) of the electric field are discontinuous, implying that the normal electric field decays by a factor of $\varepsilon_{air}/\varepsilon_{medium}$ ($\varepsilon_{air}$ and $\varepsilon_{medium}$ are the permittivity of air and medium). The focused APV beam has only the transverse electric field so that it still keeps a small spot in the medium.

The radii of the binary lens with high NA can be expressed as [18]

$$r_{2n-2} = \sqrt{(2n-2)\lambda f + (n-1)^2 \lambda^2}, \quad r_{2n-1} = \sqrt{(2n-1)\lambda f + (n-1/2)^2 \lambda^2}, \quad (1)$$

where $r_{2n-2}$ and $r_{2n-1}$ are the inner and outer radii of the $n$-th ($n$=1, 2, 3…) transparent belt of binary lens, respectively, $f$ is the focal length of meta-lens, $\lambda$ is the wavelength of the incident light. For an amplitude binary lens with the transmission of 0 and 1, its transmission function $T(r)$ is 1 where $r_{2n-2} \leq r < r_{2n-1}$ and 0 where $r_{2n-1} \leq r < r_{2n}$ (Fig. 1). In order to show the exact solution for the propagation of vector beams in a multi-layer medium after focused by binary lens, we built up a theoretical model using finite different time domain (FDTD) method. We show a simple example that two-layer (air and glass) media are involved, which is shown in Fig. 1. We assume that air and glass are individually located at $z \leq 4\lambda$ and $z > 4\lambda$ and the APV beam is incident on the binary lens at $z=0$. The binary lens having the focal length of $4\lambda$ and 10 transparent belts are modelled with penetrating through a 100nm-thickness perfect-electric-conductor film in our FDTD simulation. The interface of two media and the focal plane of binary lens share the same position. In FDTD, the glass has the refractive index $n=1.457$ at 632.8nm. Taking account for the similar focusing capacity of RP and APV beams, we just show the propagation for RP and APV beams but ignore the cases of LP and CP beams that are well-known for their poor performances in focusing [12]. The intensity profiles of RP and APV beams are considered to be the same with $P(r)=r \cdot \exp(-r^2/w^2)$ with $w$=8.5µm in FDTD.

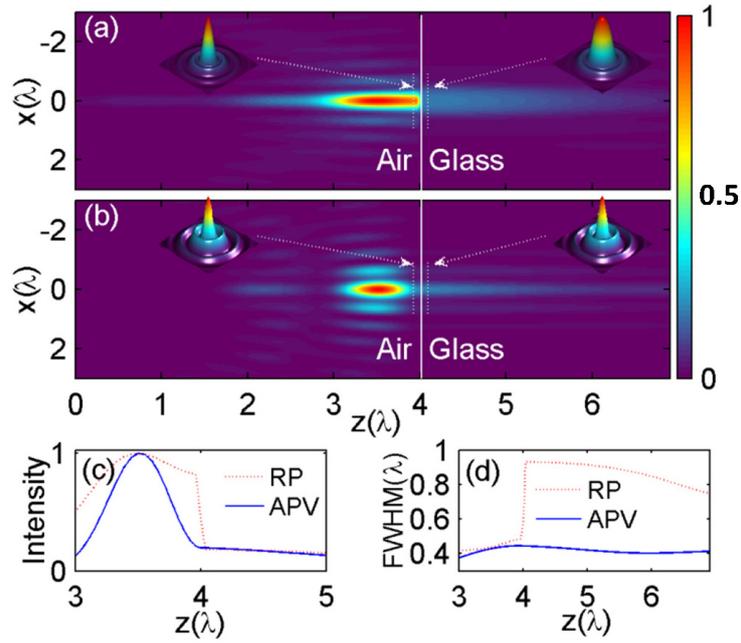

**Figure 2** (color online) Propagation of focused vector beams by a binary lens (its focal length $f=4\lambda$ and 10 transparent belts) in the air and glass (Air and glass are respectively located at $0<z\leq4\lambda$ and $z>4\lambda$ with the interface shown by the white line). (a) Intensity for RP beam. The left and right insets show the spots at $z=3.9\lambda$ and $z=4.1\lambda$, respectively. (b) Intensity for APV beam. The spots before (left) and after (right) the interface have nearly the same shape. (c) The on-axis ($x=0$) intensity profiles of RP (dot) and APV beam (solid). (d) Spot size (by FWHM) near the interface for RP (dot) and APV (solid) beams.

Figure 2(a) displays the intensity profiles of RP beam in air ($0<z\leq4\lambda$) and glass ($z>4\lambda$) after the binary lens. The focused spot before the interface has a very small spot with size of $0.48\lambda$. However, the spot size is enlarged up to $0.93\lambda$ in the glass. This abrupt increment in spot size from air to glass makes the RP beam poor in achieving a small spot. This is mainly due to the abrupt decrease of longitudinal field in glass for the boundary condition of electric field shown in Fig. 1. By contrast, the APV beam shows the exciting focusing property with super-small focused spot about $0.45\lambda$ near the interface in Fig. 2(b). Moreover, its intensity is continuous even after it goes through the same interface of air and glass, which is different from the quick fall of intensity from air to glass for the RP beam shown in Fig. 2(c). From the uniformly variant intensity for APV beam in Fig. 2(b), we can derive that the APV beam has the purely transverse electric field, which leads to the fact that the subwavelength spot maintains in the glass. In contrast with RP beam, the transversely-polarized focused APV beam also brings the stronger reflection at the air-glass interface, having the divergent strips in air as shown in Fig. 2(b). In order to evaluate the focusing efficiency of this binary lens, we use the ratio $\eta$ of the peak intensity of this focal spot to that of incident beam. The simulated $\eta$ equals 64 obtained by using FDTD under the condition that the medium is also air.

In addition, the APV beam keeps the super-resolution spot for several wavelengths in glass, which is shown in Fig. 2(b) and 2(d). In glass, the transmitted beam has the lateral size below $0.45\lambda$ at full-width of half maximum (FWHM) with the narrowest size of $0.4\lambda$ at $z=6\lambda$ shown in Fig. 2(d), and keeps its shape for about $3\lambda$ without any divergence which implies the generation of the subwavelength non-divergent beam in glass. The generation of non-divergent is due to the focal spot containing the 0-order Bessel function when APV is focused [10-12]. This small spot incident on the glass works as a source of Bessel beam, which propagates without divergence along several wavelengths in glass. However, in the homogenous medium without the interaction between the spot and interface, its focal length seems to be that of traditional spherical lens, which will be shown later.

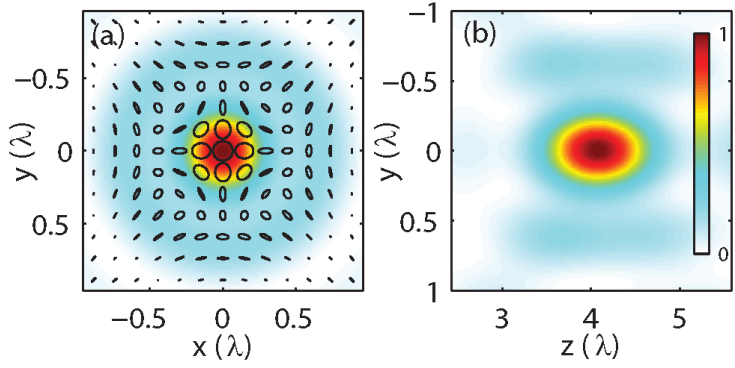

**Figure 3** (color one)Tight focusing of APV beam by a zone plates in air. (a)Intensity and polarization profiles of the focused APV at the focal plane of a FZP with $f=4\lambda$ and 16 transparent belts. The elliptical circles show the polarization of the focused beam at the focal plane. Circular polarization is only located at the centre. (b)Side-view ($y$-$z$ plane) intensity profile

The APV beam also has the good focusing performance in a single homogenous medium (e.g. air). The tight focusing of a vector beam by a high NA binary lens in a single medium can be approximated by using the vectorial Rayleigh-Sommerfeld diffraction [16, 19]. For simplicity, we investigate the case in air, which implies that the medium in Fig. 1 is air. For an APV beam with its electric field $\mathbf{E}_i(r,\varphi,0)=P(r)e^{i\varphi}\mathbf{e}_\varphi$ [10] at the plane $z=0$ where the binary lens is located, the electric field of transmitted light after the binary lens can be expressed by

$$\mathbf{E}(\rho,\phi,z) = \frac{1}{2\pi} \int_0^\infty \int_0^{2\pi} \begin{bmatrix} E_{ix}(r,\varphi,0)\mathbf{e}_x \\ E_{iy}(r,\varphi,0)\mathbf{e}_y \\ 0\mathbf{e}_z \end{bmatrix} T(r)\left(ik - \frac{1}{R}\right)\frac{z}{R^2} \cdot \exp(ikR) r\,dr\,d\varphi , \quad (2)$$

where $R^2=r^2+\rho^2+z^2-2r\rho\cos(\varphi-\phi)$. The transmitted light only has the transverse field component according to Eq. (2). For a high NA binary lens, its focal length $f$ is short in contrast with its maximum radius $r_{max}$, e.g. $f=4\lambda$ with $r_{max}=19.1\lambda$ used here.

Although an azimuthally polarized beam has the intensity profile obeying the Bessel-Gaussian distribution [20], we assume that the APV beam has a uniform intensity distribution except the singularity at the center in our simulation for the convenience of comparing with other polarizations. Here, we show the simulated intensity profiles in the focal region of binary lens in Fig. 3. Figure 3(a) shows the intensity profile and polarization pattern at the focal plane. The spot has the size of $0.455\lambda$, which is below the Rayleigh diffraction limitation of $0.526\lambda$ (=$0.515\lambda/0.95$ by FWHM). This is due to introducing the vortical phase into the azimuthally polarized beam for the constructive interference [13].

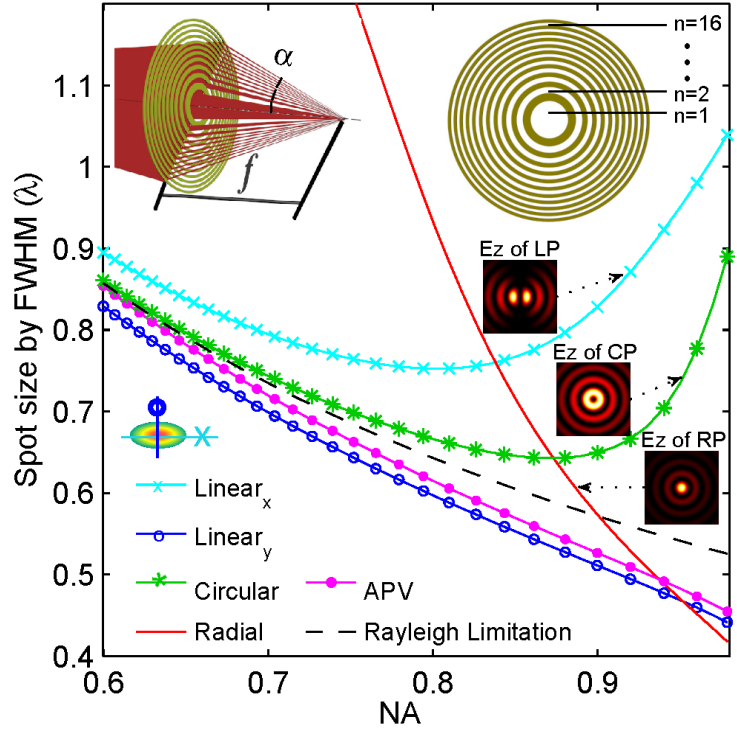

**Figure 4** (color online) Focused spot size of the vector beams by a binary lens with various NA from 0.6 to 0.98. The NA of lens is changed by adjusting its focal length $f$ (left inset) with 16 transparent belts (right inset), resulting in the change of maximum convergent angle α that is relative with NA(=sinα). The linearly-polarized light has an elliptical focusing spot with its different size in $x$ (blue circle) and $y$ (cross) direction. The pseudo-color pictures show the intensity profiles of longitudinal field $E_z$ by focusing the linearly (LP), circularly (CP) and radially (RP) polarized beams with a binary lens.

In addition, the vortical phase changes the polarization pattern at the focal plane as shown in Fig. 3(a). The circular polarization is located at the center of spot surrounded by the elliptical polarization with slowly various elliptical degree. The circularly polarization at the center can be easily verified by setting $\rho=0$ in Eq. 2, resulting that the variable $R$ is independent on the angle coordinate $\varphi$. The integral over $\varphi$ can be extracted from the double integral, leaving the same integral over $r$ for both of $x$ and $y$ component fields. Finally, these two integrals over $\varphi$ are $\pi$ and $\pi \cdot i$, respectively.

As depicted in Fig. 3(b), the APV beam has a depth of focus (DOF) of about $1.1\lambda$ when focused by a high NA binary lens. This is smaller than the DOF ($2.1\lambda=2\lambda/NA^2$) of traditional lens [21]. The short DOF implies the high axial resolution in the imaging system. When it is used in a focusing system, e.g. optical lithography, the short DOF requires more precise control of axial position, which is not difficult nowadays. In contrast with the short DOF in focusing system, one will pursue the small spot.

Figure 4 displays the focused spot size at the focal plane when the vector beams illuminate the binary lens with different NA. The LP ($x$-polarization used here) beam has the focused spot with elliptical shape because the saddle $E_z$ leads the different FWHMs in $x$ and $y$ direction as shown in Fig. 4. Another well-known CP beam also has the bad focusing properties for the binary lens in a wide range of NA. Like the changing tendency of spot size in $x$-direction for LP beam, the spot size for CP beam also undergoes a valley, upper the Rayleigh limitation, around the NA=0.9. The valleys for LP and CP beams mainly depend on the intensity patterns from their longitudinal electric fields, which are shown in Fig. 4. When the NA of the binary lens increases in the range from 0.8 to 1, their longitudinal electric fields will increase abruptly and dominate the focal region, resulting in the increment of spot size. However, the RP beam has the sharp decrement in spot size because its longitudinal electric field ($E_z$) is a hotspot, as shown in Fig. 4. It is worthy to note that the spot for RP beam is beyond the Rayleigh limitation only for a binary lens with NA>0.9.

Fortunately, the APV beam has no longitudinal electric field in the focal region and shows the perfect focusing beyond the Rayleigh limitation for the binary lens with NA ranged in [0.6, 1], which behaves better than RP, LP and CP beams. Although the RP beam has the smaller spot than APV beam when NA>0.94, the difference between their focused spots is insignificant, i.e. two spots at NA=0.98 only has the difference of 0.03λ. Considering the APV beam's focusing performance in generating a super-resolution spot by a zone plates lens with a wide range of NA, we can call the APV beam as "super-resolution beam". The "super-resolution" means that the focused spot size (not the lateral size of beam) after focusing the beam by a lens (without any additional amplitude, phase and polarization modulation) is below the Rayleigh limitation. The RP beam can be taken as the 'super-resolution' beam only when the NA of focusing lens is larger than 0.9 as shown in Fig. 4. To our knowledge, the APV beam is the only one in vector beams that can be considered as the "super-resolution beam" in a wide range of NA, which may be further integrated with advanced lens such as metalens [22], superlens [23] and super-oscillatory lens [15, 24] so as to further improve imaging capabilities.

In summary, we have demonstrated the focusing properties of vector beams by a high NA binary lens with one medium and two-layer medium located in the focal region. The APV beam has a focused spot beyond the Rayleigh limitation for the binary lens with a wide range of NA, which makes it become the so-called "super-resolution beam". In addition, we also show the generation of a subwavelength non-divergent beam in glass after the focused APV beam passes through the interface of air and glass. This provides new opportunities to nano-photonics for manipulating light in the subwavelength scale for vectorial focusing and imaging.